# The Mitochondrial Genome of *Cathaya argyrophylla* Reaches 18.99 Mb: Analysis of Super-Large Mitochondrial Genomes in Pinaceae

Kerui Huang[1], Wenbo Xu[2], Haoliang Hu[1], Xiaolong Jiang[3], Lei Sun[4], Wenyan Zhao[1], Binbin Long[1], Shaogang Fan[1], Zhibo Zhou[1], Ping Mo[1], Xiaocheng Jiang[5], Jianhong Tian[4], Aihua Deng[1], Peng Xie[1], Yun Wang[1]

[1]Key Laboratory of Agricultural Products Processing and Food Safety in Hunan Higher Education, Science and Technology Innovation Team for Efficient Agricultural Production and Deep Processing at General University in Hunan Province, Hunan University of Arts and Science, Changde 415000, China.

[2]Chinese Medicine Research Institute of Beijing Tcmages Pharmaceutical Co., Ltd., Beijing 101301, China

[3]College of Forestry, Central South University of Forestry and Technology, Changsha 415000, China

[4]Key Laboratory of Research and Utilization of Ethnomedicinal Plant Resources of Hunan Province, College of Biological and Food Engineering, Huaihua University, Huaihua 418000, China.

[5]College of Life Sciences, Hunan Normal University, Changsha 410081, China.

## Abstract

Mitochondrial genomes in the Pinaceae family are notable for their large size and structural complexity. In this study, we sequenced and analyzed the mitochondrial genome of *Cathaya argyrophylla*, an endangered and endemic Pinaceae species, uncovering a genome size of 18.99 Mb—the largest mitochondrial genome reported to date. To investigate the mechanisms behind this exceptional size, we conducted comparative analyses with other Pinaceae species possessing both large and small mitochondrial genomes, as well as with other gymnosperms. We focused on repeat sequences, transposable element activity, RNA editing events, chloroplast-derived sequence transfers (mtpts), and sequence homology with nuclear genomes. Our findings indicate that while *Cathaya argyrophylla* and other extremely large Pinaceae mitochondrial genomes contain substantial amounts of repeat sequences and show increased activity of LINEs and LTR retrotransposons, these factors alone do not fully account for the genome expansion. Notably, we observed a significant incorporation of chloroplast-derived sequences in *Cathaya argyrophylla* and other large mitochondrial genomes, suggesting that extensive plastid-to-mitochondrial DNA transfer may play a crucial role in genome enlargement. Additionally, large mitochondrial genomes exhibited distinct patterns of RNA editing and limited similarity with nuclear genomes compared to smaller genomes. These results suggest that the massive mitochondrial genomes in Pinaceae are likely the result of multiple contributing factors, including repeat sequences, transposon activity, and extensive plastid sequence incorporation. Our study enhances the understanding of mitochondrial genome evolution in plants and provides valuable genetic information for the conservation and study of *Cathaya argyrophylla*.

## Introduction

Mitochondria play a crucial role in energy metabolism within plant cells, and their genomes are essential for studies on plant evolution, physiological functions, and genetic diversity [1-4]. While the mitochondrial genomes of animals and fungi are relatively simple in structure, plant mitochondrial genomes are known for their significant variation in size, structural complexity, and frequent recombination [2,5-8]. Research on plant mitochondrial genomes helps to elucidate plant cell evolution, the origins and differentiation of species, the mechanisms behind genomic structural diversity, as well as the interaction and co-evolution between the mitochondrial and nuclear genomes [2,4,9,10].

In plant mitochondrial genomes, the substantial variation in genome size is particularly striking. The mitochondrial genome of plants ranges from 66 kb in the parasitic plant *Viscum scurruloideum* to 11.7 Mb in the Siberian larch *(Larix sibirica*), with considerable differences even between closely related species [11-17]. This provides a unique opportunity to investigate the evolutionary mechanisms underlying plant mitochondrial genome diversity. Among plants with exceptionally large mitochondrial genomes, Pinaceae family have garnered attention due to the massive mitochondrial genomes reported in several species [5,9,17]. For instance, Larix sibirica possesses the largest known mitochondrial genome at 11.7 Mb [17], and *Picea sitchensis* also has a notably large mitochondrial genome at 5.5Mb [9]. Although initial reports exist, the commonalities and mechanisms behind the enormous size of Pinaceae mitochondrial genomes remain largely unexplored.

*Cathaya argyrophylla*, as an ancient member of the Pinaceae family and endemic to China, is renowned for its extreme endangerment and rarity, earning the nickname "Panda of the plant kingdom" due to its special evolutionary status and research value [18-20]. In the early stage of this study, we sequenced and assembled its mitochondrial genome, revealing that, similar to other Pinaceae species with massive mitochondrial genomes, *Cathaya argyrophylla* also possesses an exceptionally large genome at 18.99 Mb, it is the largest mitochondrial genome reported to date, breaking previous records. However, why do Pinaceae species, including *Cathaya argyrophylla*, harbor such enormous mitochondrial genomes? What are the common characteristics of these large genomes, and how do they differ from or relate to the smaller mitochondrial genomes of other Pinaceae species and gymnosperms? Addressing these questions is crucial for understanding the structure, function, and evolution of plant mitochondrial genomes.

Previous studies have suggested that factors such as repeat sequences, transposon activity, RNA editing events, and the transfer of genomic fragments between organelles may play key roles in the amplification and complexity of plant mitochondrial genomes [1,2,4,13,15,21]. However, the commonalities of these factors and their potential contribution to the genome size expansion in Pinaceae mitochondrial genomes remain largely unknown. Furthermore, the extent to which genomic exchanges between the mitochondrial genome, nuclear genome, and chloroplast genome in Pinaceae species are linked to the large mitochondrial genome sizes is also a topic that warrants further investigation.

To address the aforementioned scientific questions, this study conducted a comprehensive analysis of the mitochondrial genome of *Cathaya argyrophylla*, focusing on several key aspects: the quantity, coverage, types patterns of repeat sequences; the patterns of transposable elements, particularly LINEs and LTR elements; the number and patterns of RNA editing events; the quantity, length, and coverage of chloroplast-derived fragments (mtpts); and the sequence homology between the mitochondrial genome and the nuclear genome. We also performed a systematic comparison between *Cathaya argyrophylla* and other Pinaceae species with large mitochondrial genomes as well as species with smaller mitochondrial genomes (for detailed definitions of large and small genomes, refer to the Materials and Methods section), along with other gymnosperms, in order to uncover common features and potential mechanisms behind the formation of large mitochondrial genomes. Through this study, we aim to elucidate the uniqueness of large Pinaceae mitochondrial genomes and explore the relationship between genome size and factors such as repeat sequences, transposable element activity, and organelle-to-organelle sequence exchange. Our findings will provide new perspectives and evidence for understanding the characteristics and evolution of plant mitochondrial genomes. Furthermore, the in-depth study of *Cathaya argyrophylla* will offer essential genetic foundations for its conservation and potential utilization.

## Materials and Methods

### Plant Sample Collection and DNA Extraction

We collected 50 g of root tissue from cultivated *Cathaya argyrophylla* in Xinning County, Hunan Province (26°33′N, 110°36′E). The roots were washed with distilled water and immediately flash-frozen in liquid nitrogen for preservation.

Genomic DNA was extracted from the above-mentioned root tissue using the CTAB method. Approximately 0.5 g of fresh tissue was ground in liquid nitrogedn using a tissue grinder (Tissuelyser, Shanghai Jingxin) and mixed with 20 mL of CTAB extraction buffer (Biosharp, BS950-500). The mixture was incubated at 50°C for 1.5 hours in a programmable metal bath (OSE-DB-02, TIANGEN), with gentle inversion every 20 minutes using a vortex mixer (Vortex-5, Qilinbeier). After incubation, the lysate was cooled to room temperature and centrifuged at 10,000 g for 10 minutes in a centrifuge (H1850, Xiangyi). The supernatant was extracted twice, first with phenol/chloroform/isoamyl alcohol (25:24:1, BioFroxx) and then with chloroform/isoamyl alcohol (24:1, BioFroxx). DNA was precipitated using 0.8 volumes of isopropanol (Sinopharm, 80109218), washed twice with 75% ethanol (Sinopharm, 10009218), air-dried, and dissolved in 100 μL of elution buffer (Qiagen, 19086). RNA contamination was removed by treating the DNA with RNase A (Solarbio, 9001-99-4) at 37°C for 30 minutes. The extracted DNA was quantified using a NanoDrop spectrophotometer (NanoDrop One, Thermo Scientific) and a Qubit fluorometer (Qubit 3.0, Thermo Scientific), with quality verified by agarose gel electrophoresis (Biowest, 111860) stained with GoldView nucleic acid dye (Biosharp, BS357A).

### Mitochondrial Genome Sequencing

Genomic DNA was extracted and purified using a CTAB-based method, followed by quality control using a Qubit 3.0 Fluorometer (Thermo Scientific) and 1% agarose gel electrophoresis. For long-read sequencing, libraries were prepared using the Oxford Nanopore Technologies (ONT) Ligation Sequencing Kit (SQK-LSK110) according to the manufacturer's instructions. A total of 2.5 μg of high-quality DNA was used, and end-repair was performed using the NEBNext FFPE DNA repair kit (NEB, M6630L) with 20°C incubation for 10 minutes, followed by 65°C for 10 minutes. Sequencing adapters were ligated using the NEBNext Quick T4 DNA Ligase (NEB, E6056S) at 25°C for 10 minutes. The libraries were then purified using AMPure XP beads (Beckman Coulter, A63881) and eluted in 25 μL of Elution Buffer (Qiagen, 19086). The prepared libraries were loaded onto a PromethION sequencer (Oxford Nanopore Technologies) and sequenced for 48-72 hours.

For short-read sequencing, libraries were prepared using the Plus DNA Library Prep Kit (MGI, NDM627) on the DNBSEQ-T7 platform. A total of 1 μg of genomic DNA was fragmented using the Covaris system or by enzymatic digestion to achieve fragments of 200-400 bp. The fragments were end-repaired and A-tailed, followed by adapter ligation at 20°C for 30 minutes. The libraries were PCR-amplified, purified using AMPure XP beads, and circularized. The final circularized libraries were loaded onto the DNBSEQ platform for high-throughput sequencing.

**Mitochondrial Genome Assembly and Annotation**

The mitochondrial genome contig fragments of *Cathaya argyrophylla* were first assembled using Flye with long-read sequencing data [22]. Subsequently, both the long- and short-read sequences were aligned to these fragments using BWA [23]. The aligned reads were then assembled into a complete mitochondrial genome utilizing Unicycler [24]. Visualization and export of the final assembled genome were performed with Bandage [25]. The mitochondrial genome was annotated to identify protein-coding genes, tRNAs, and rRNAs using IPMGA, tRNAscan-SE, and BLASTn, respectively. *Pinus taeda* (NC037304) and *Liriodendron tulipifera* (NC021152) were used as reference genomes. Manual corrections were made in Apollo [26], and the final annotation files were submitted to the NCBI database under accession numbers PP764533 to PP764541. Additionally, the raw data from both second- and third-generation sequencing have been submitted to NCBI under the following identifiers: BioProject: PRJNA1105731, BioSample: SAMN41108760, and SRA: SRR28842127 (third-generation) and SRR28842128 (second-generation).

**Preparation of Mitochondrial Genomes for Comparison and Classification of Large and Small Genomes**

In this study, we selected the mitochondrial genomes of 11 gymnosperm species, including the mitochondrial genome of *Cathaya argyrophylla*, for subsequent comparative analyses. These species were chosen based on the availability of complete mitochondrial genome assemblies from NCBI, which included *Abies koreana* (NC071216), *Cathaya argyrophylla* (PP764533-PP764541), *Cycas taitungensis* (AP009381), *Ginkgo biloba* (KM672373), *Larix sibirica* (MT797187-MT797195), *Picea sitchensis* (MK697696-MK697708), *Pinus taeda* (MF991879), *Platycladus orientalis* (OL703044-OL703045), *Taxus cuspidata* (MN593023), *Thuja sutchuenensis* (ON603305-ON603308), and *Welwitschia mirabilis* (KT313400). These sequences were downloaded for subsequent analysis.

For the purposes of this study, mitochondrial genomes less than 2M in size were classified as "small," genomes greater than or equal to 2M as "large," and genomes greater than or equal to 5M as "extremely large."

**Synteny Analysis**

We first used BLASTn to identify homologous fragments between the mitochondrial genome of *Cathaya argyrophylla* and those of 10 other gymnosperms. Only fragments larger than 200 bp were retained. These homologous fragments were then processed into a format recognizable by MCscanX [27] using a custom Python script, based on their positions. MCscanX was subsequently employed to plot multiple synteny plots of *Cathaya argyrophylla* and the 10 other gymnosperms, based on the syntenic blocks formed by the aforementioned homologous fragments.

**Repeat Sequence Analysis**

RepeatModeler (https://www.repeatmasker.org/) was used to identify repeat elements in the

mitochondrial genome assembly. TEclass online service [28] was utilized to classify unknown repeat elements from the de novo repeat element library generated by RepeatModeler. A custom Python script was applied to parse RepeatMasker results based on the RepBase classification.

**RNA Editing Analysis**

To ensure accuracy in RNA editing analysis, we used transcriptome data from 10 samples of the same individual of *Cathaya argyrophylla* (unpublished data). The data were converted into BAM files using Samtools, and Samtools mpileup was used to align them to the mitochondrial genome's protein-coding genes of *Cathaya argyrophylla*. A custom Python script was employed to detect RNA editing sites. An RNA editing site was defined if a particular edit occurred more than 60% of the time, and these were statistically recorded. RNA editing site data for other species were obtained from previously published studies.

**MTPT Sequence Transfer Analysis**

To identify potential homologous sequences transferred between the plastid and mitochondrial genomes of *Cathaya argyrophylla* and other gymnosperm species, we conducted a BLASTn [29] comparison between the two organelles using the following parameter settings: E-value = 1e-5. Visualization of the BLAST results was performed using TBtools [30]. A custom Python script was employed to determine whether the homologous sequences contained chloroplast genes and to assess the completeness of these genes.

The chloroplast genome of *Cathaya argyrophylla* used in this analysis was obtained from our previous study (NCBI accession: OL790355). The chloroplast genomes of the other species were as follows: *Abies koreana* (NC026892), *Cycas taitungensis* (NC009618), *Ginkgo biloba* (MN443423), *Larix sibirica* (NC036811), *Picea sitchensis* (NC011152), *Pinus taeda* (KY964286), *Platycladus orientalis* (KX832626), *Taxus cuspidata* (NC041498), *Thuja sutchuenensis* (NC042176), and *Welwitschia mirabilis* (EU342371). The mitochondrial genomes were mentioned in the " Preparation of Mitochondrial Genomes for Comparison and Classification of Large and Small Genomes" section and will not be reiterated here.

**NUMTs Sequence Transfer Analysis**

This study analyzed the sequence exchange between the mitochondrial and nuclear genomes (NUMTs) in gymnosperms and angiosperms. For angiosperms, we downloaded the complete nuclear genome sequences of *Arabidopsis thaliana* (GCF_000001735.4), *Asparagus officinalis* (GCF_001876935.1), *Glycine max* (GCF_000004515.6), *Nicotiana tabacum* (GCF_000715135.1), *Oryza sativa* (GCF_034140825.1), and *Zea mays* (GCF_902167145.1). The first three species represented dicotyledons, while the latter three represented monocotyledons. For gymnosperms, we downloaded the complete nuclear genome sequences of *Pinus taeda* (GCA_000404065.3), *Larix sibirica* (GCA_004151065.3), *Picea sitchensis* (GCA_010110895.2), and *Cycas panzhihuaensis* (GCA_023213395.1).

We used the complete nuclear genome sequences of the aforementioned species as indexes. Then, BLASTn was employed to align the mitochondrial genomes of Pinaceae species and gymnosperms to these nuclear genome indexes to identify homologous fragments. The BLASTn parameter settings were: E-value = 1e-5, and only fragments longer than 100 bp were retained as NUMTs. Subsequent analyses were performed using a custom Python script.

## Results

### Assembly and Annotation of the *Cathaya argyrophylla* Mitochondrial Genome

Through sequencing, assembly, and annotation of the mitochondrial genome of *Cathaya argyrophylla*, we identified its primary structure as a multi-branched configuration. After excluding repeat regions using Nanopore data, we obtained six linear contigs and three circular contigs (Figure 1), with a total length of 18,990,836 bp and a GC content of 44.06%. The assembly results indicate that *Cathaya argyrophylla* possesses the largest mitochondrial genome reported to date (Table 1). The mitochondrial genome annotation of *Cathaya argyrophylla* revealed 40 unique protein-coding genes, including 24 unique mitochondrial core genes and 16 non-core genes, 32 tRNA genes (with 25 tRNA genes being multi-copy), and 3 rRNA genes (with all three being multi-copy). The core genes included 5 ATP synthase genes (*atp1*, *atp4*, *atp6*, *atp8*, and *atp9*); 9 NADH dehydrogenase genes (*nad1*, *nad2*, *nad3*, *nad4*, *nad4L*, *nad5*, *nad6*, *nad7*, and *nad9*); 4 cytochrome c biogenesis genes (*ccmB*, *ccmC*, *ccmFC*, and ccmFN*);* 3 cytochrome c oxidase genes (*cox1*, *cox2*, and *cox3*); 1 membrane transporter gene (*mttB*); 1 maturase gene (*matR*); and 1 ubiquinol-cytochrome c reductase gene (cob). The non-core genes consisted of 3 large ribosomal subunit genes (rpl2, rpl5, and rpl16); 11 small ribosomal subunit genes (*rps1*, *rps2*, *rps3*, *rps4*, *rps7*, *rps10*, *rps11*, *rps12, rps13*, *rps14*, and *rps19*); and 2 succinate dehydrogenase genes (*sdh3* and *sdh4*).

Table 1: Comparison of Mitochondrial Genome Sizes between *Cathaya argyrophylla* and Other Plant Species

| Species | Mitochondrial Genome Size (Mbp) | Group | Source (NCBI/Reference) |
|---|---|---|---|
| *Cathaya argyrophylla* | 18.99 | extremely large | This study （PP764533-PP764541） |
| *Larix sibirica* | 11.69 | extremely large | MT797187-MT797195 |
| *Picea smithiana* | 6.17 | extremely large | Kan et al. 2021[31] |
| *Picea sitchensis* | 5.52 | extremely large | MK697696-MK697708 |
| *Cedrus deodara* | 5.13 | extremely large | Kan et al. 2021[31] |
| *Pinus armandii* | 3.85 | large | Kan et al. 2021[31] |
| *Taiwania cryptomerioides* | 2.80 | large | Kan et al. 2021[31] |
| *Platycladus orientalis* | 2.46 | large | OL703044-OL703045 |
| *Thuja_sutchuenensis* | 2.23 | large | ON603305-ON603308 |
| *Cunninghamia lanceolata* | 1.89 | small | Kan et al. 2021[31] |
| *Sciadopitys verticillata* | 1.87 | small | Kan et al. 2021[31] |
| *Metasequoia glyptostroboides* | 1.77 | small | Kan et al. 2021[31] |
| *Araucaria cunninghamii* | 1.54 | small | Kan et al. 2021 |
| *Podocarpus macrophyllus* | 1.39 | small | Kan et al. 2021[31] |
| *Abies firma* | 1.33 | small | Kan et al. 2021[31] |
| *Pinus taeda* | 1.19 | small | MF991879 |
| *Abies_koreana* | 1.17 | small | NC_071216 |
| *Gnetum montanum* | 1.13 | small | Kan et al. 2021[31] |
| *Welwitschia mirabilis* | 0.98 | small | KT313400 |
| *Zamia furfuracea* | 0.62 | small | Kan et al. 2021[31] |
| *Liriodendron tulipifera* | 0.55 | small | Kan et al. 2021[31] |
| *Oryza sativa* | 0.49 | small | Kan et al. 2021[31] |
| *Taxus cuspidata* | 0.47 | small | MN593023 |
| *Ephedra przewalskii* | 0.45 | small | Kan et al. 2021[31] |
| *Cycas taitungensis* | 0.41 | small | AP009381 |
| *Cycas revoluta* | 0.41 | small | Kan et al. 2021[31] |
| *Arabidopsis thaliana* | 0.37 | small | Kan et al. 2021[31] |
| *Ginkgo biloba* | 0.35 | small | KM672373 |
| *Cephalotaxus sinensis* | 0.34 | small | Kan et al. 2021[31] |

Note: This study primarily includes species with fully assembled mitochondrial genomes and available NCBI accession numbers for comparative analysis. In the table, species with NCBI accession numbers were selected for comparative analysis.

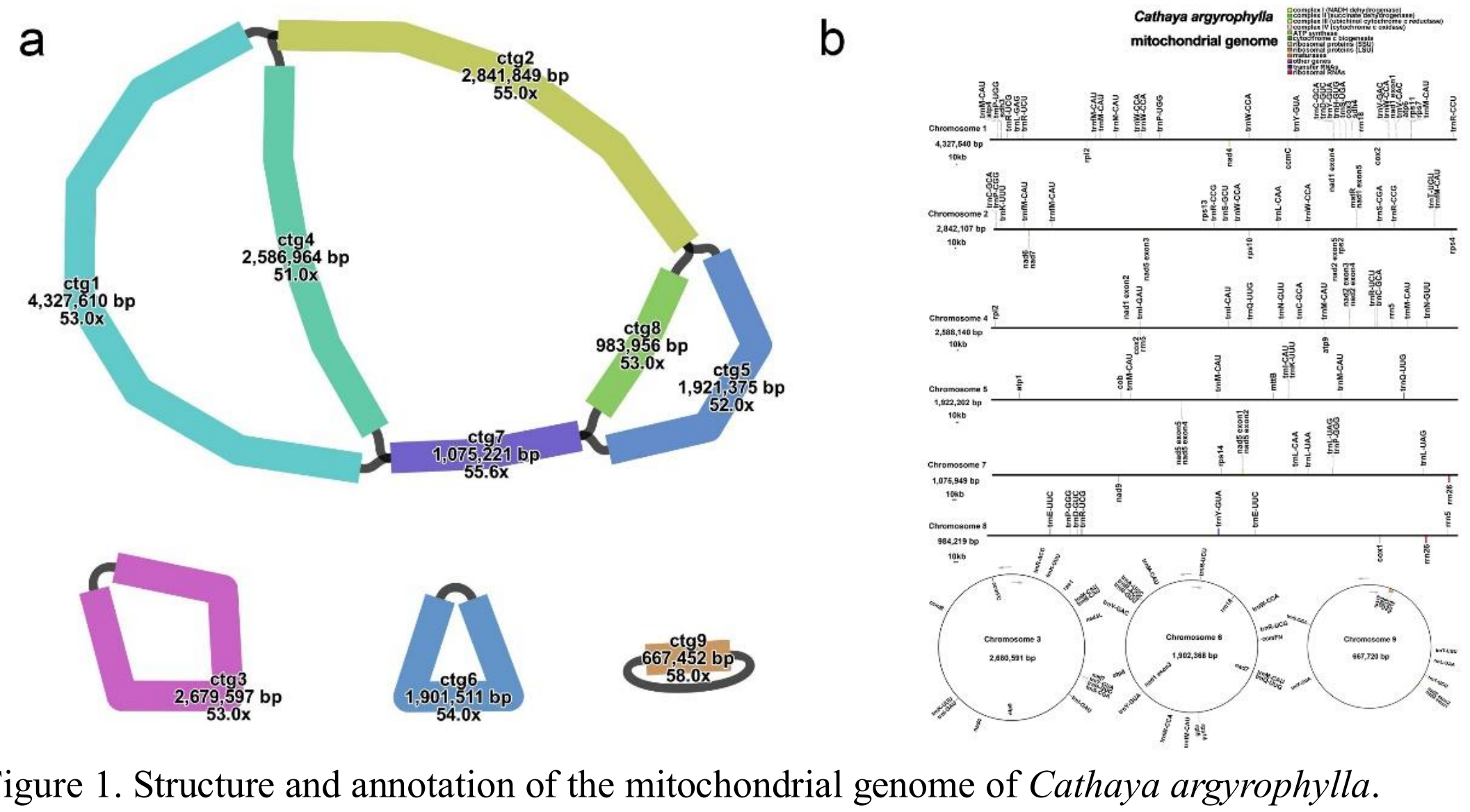

Figure 1. Structure and annotation of the mitochondrial genome of *Cathaya argyrophylla*.

Note: (a) The structure of *Cathaya argyrophylla* mitochondrial genome, illustrating its multi-branched configuration, including six linear contigs and three circular contigs. (b) Gene annotation information of *Cathaya argyrophylla*, highlighting protein-coding genes, tRNAs, rRNAs, and other functional elements.

**Comparative Synteny Analysis Among Pinaceae Mitochondrial Genomes**

To explore the homology between the mitochondrial genome sequences of *Cathaya argyrophylla* and those of Pinaceae plants, we conducted a synteny analysis between the mitochondrial genomes of *Cathaya argyrophylla* and other gymnosperms (Figure 2). We considered syntenic blocks as arrangements of homologous fragments with lengths greater than or equal to 200 bp. The results revealed that *Cathaya argyrophylla* shares numerous syntenic blocks with Pinaceae species, and these blocks often display significant rearrangements. Furthermore, *Cathaya argyrophylla* has more syntenic blocks with species possessing larger mitochondrial genomes, though the number of these blocks decreases as phylogenetic distance increases. For example, *Larix sibirica*, which has the largest mitochondrial genome after *Cathaya argyrophylla*, shares 83 syntenic blocks with *Cathaya argyrophylla*. In contrast, *Picea sitchensis*, which is more closely related to *Cathaya argyrophylla* and has a mitochondrial genome nearly half the size of *Larix sibirica*, shares 84 syntenic blocks. Additionally, while the mitochondrial genomes of *Pinus taeda* and *Abies koreana* are similar in size and their mitochondrial genomes are both small, *Pinus taeda*, which is more closely related to *Cathaya argyrophylla*, shares 43 syntenic blocks, whereas *Abies koreana*, which is more distantly related, shares only 18 blocks. For non-Pinaceae species, the number of syntenic blocks shared with *Cathaya argyrophylla* is significantly lower, all being fewer than or equal to 10. This suggests that closer phylogenetic relationships correspond to higher mitochondrial genome sequence homology, confirming the relatedness among Pinaceae species. Additionally, species with larger mitochondrial genomes in Pinaceae appear to share certain characteristics that result in a higher number of syntenic blocks, indicating that mitochondrial genome size may influence syntenic relationships within Pinaceae.

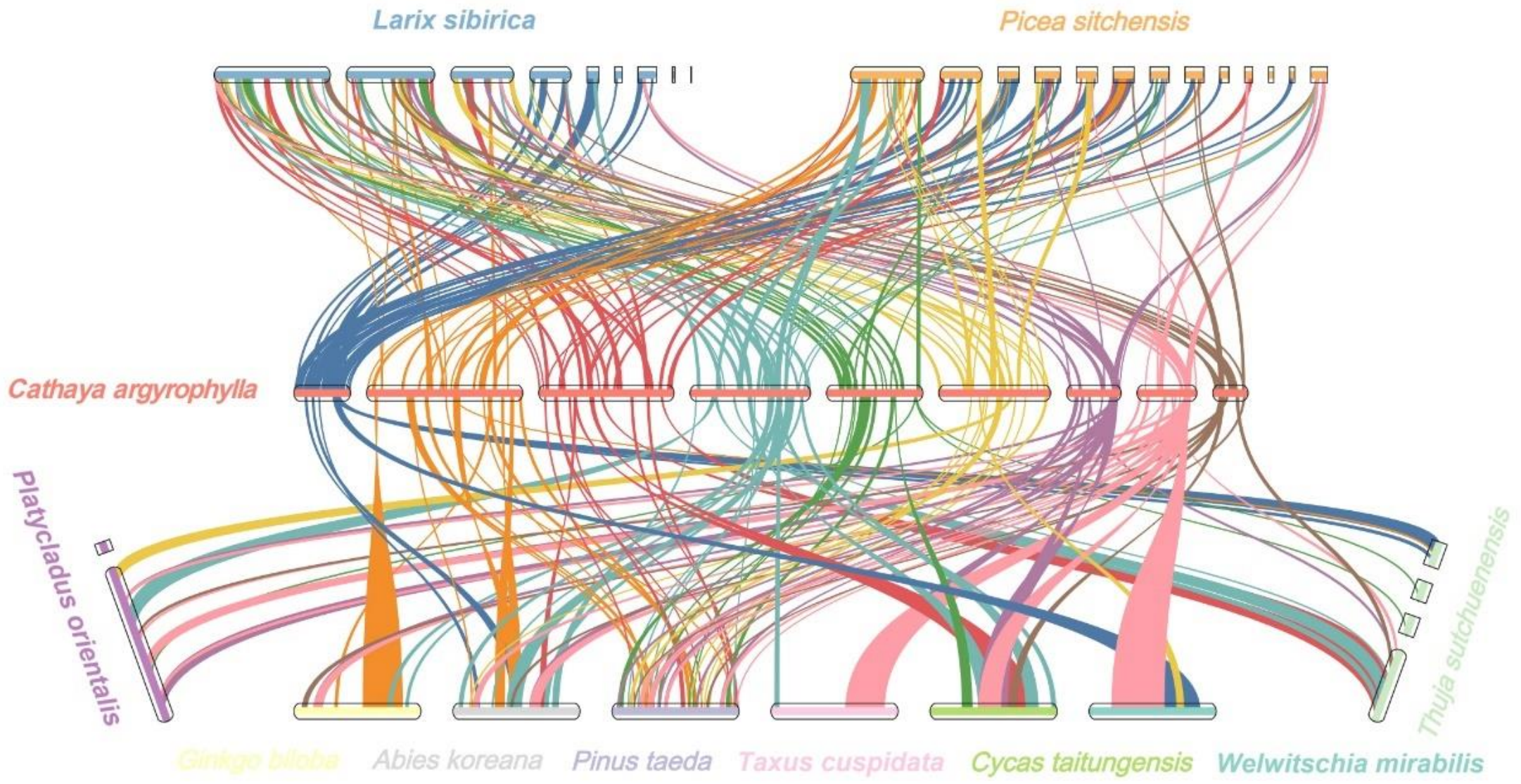

Figure 2. Synteny analysis between *Cathaya argyrophylla* and other gymnosperms, showing homologous blocks and structural conservation across mitochondrial genomes.

**Relationship Between Repeat Sequences and Mitochondrial Genome Size in Pinaceae**

Analysis of repeat sequences shows that *Cathaya argyrophylla*, similar to *Larix sibirica*, which also has an extremely large mitochondrial genome, possesses the longest total length of repeat sequences, as well as the longest total length of its genome covered (masked) by these repeats (Figure 3 and Figure 4). However, the proportion of *Cathaya argyrophylla*'s mitochondrial genome covered by repeat sequences is not the highest. In fact, *Larix sibirica* has an even lower coverage, with the highest coverage in *Cathaya argyrophylla* being only 14.78%, and *Larix sibirica* at 9.08%. In contrast, small mitochondrial genomes, such as those of Thuja sutchuenensis (31.10%), *Abies koreana* (18.46%), and *Cycas taitungensis* (17.88%), exhibit the highest repeat sequence coverage, all surpassing that of *Cathaya argyrophylla* and *Larix sibirica*. This suggests that large mitochondrial genomes are not predominantly composed of repeat sequences. Interestingly, in species with extremely large mitochondrial genomes (such as *Cathaya argyrophylla*, *Larix sibirica*, and *Picea sitchensis*), the coverage of repeat sequences increases with genome size, indicating that among extremely large mitochondrial genomes, larger mitochondrial genomes are more prone to developing repeat sequences. This pattern is not observed in other mitochondrial genomes, where some of the smallest genomes show the highest repeat sequence coverage, highlighting significant differences in sequence composition between extremely large and smaller mitochondrial genomes.

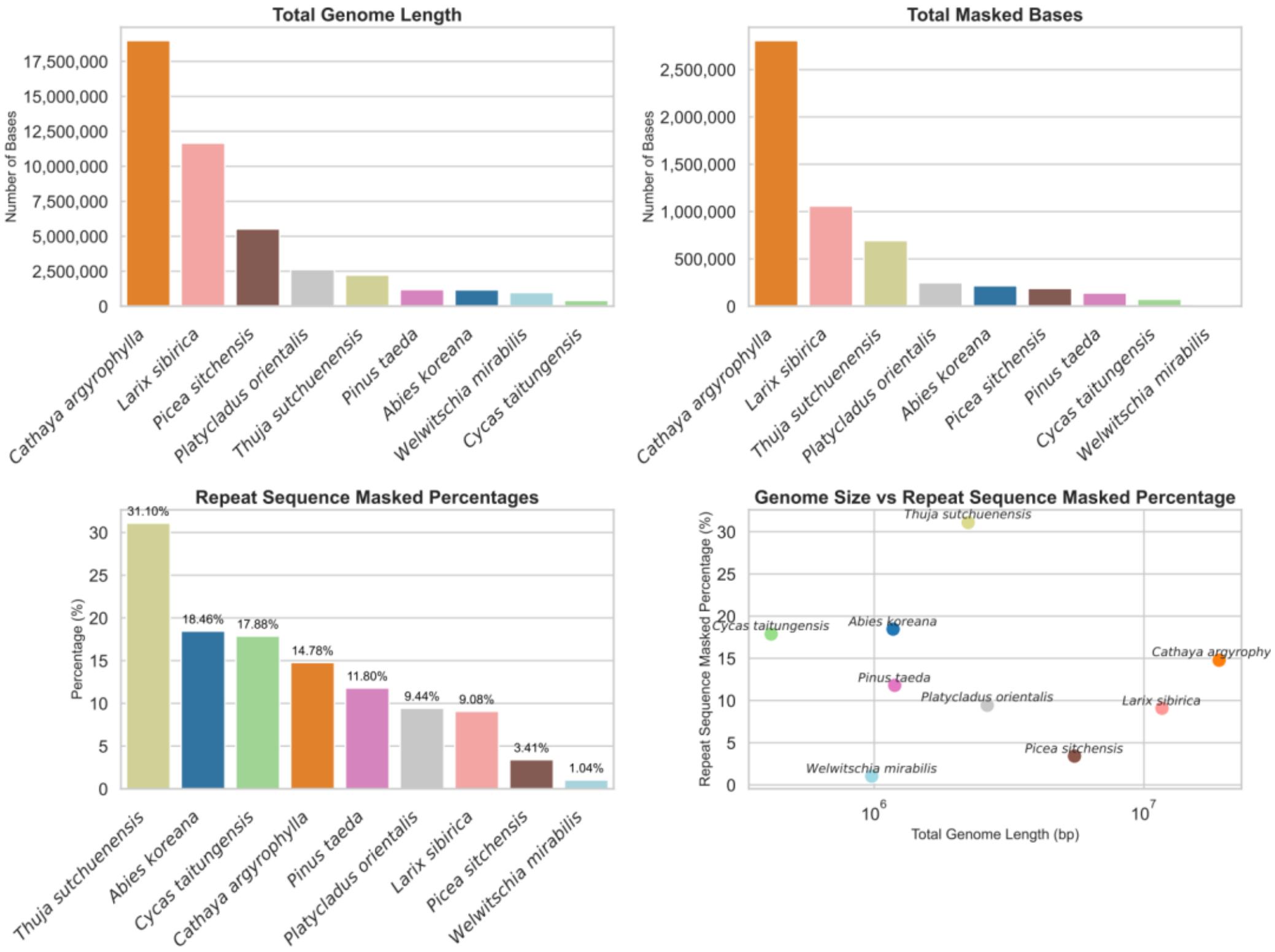

Figure 3. Comparative analysis of repeat sequences between *Cathaya argyrophylla* and other gymnosperms.

Note: Ginkgo biloba and Taxus cuspidata had no detectable repeat sequences, and thus are not included in the figure.

In terms of repeat sequence composition, only five types were identified in the mitochondrial genomes of all species: LINEs, LTR elements, low-complexity sequences, simple repeats, and unclassified sequences. Among these, unclassified sequences were the most common, while the autonomous retrotransposon LINEs were the rarest, and entirely absent in smaller mitochondrial genomes. However, in extremely large genomes, LINEs began to appear in small quantities in Larix sibirica, and their proportion relative to the total genome length was higher in *Cathaya argyrophylla*, which possesses the largest mitochondrial genome. This suggests that the activity of LINEs increases as mitochondrial genome size grows among extremely large mitochondrial genomes. A similar trend was observed for another autonomous retrotransposon, LTR elements, which, though generally rare, increased in proportion as mitochondrial genome size grew among species with extremely large mitochondrial genomes. This indicates that among extremely large mitochondrial genomes, larger mitochondrial genomes tend to exhibit higher transposon activity, a phenomenon not observed in smaller genomes. This suggests that extremely large mitochondrial genomes may have experienced more frequent high-activity events involving LINEs and LTR elements, and that genome size is likely related to this activity. However, an exception was found in *Abies koreana*, a species with a small mitochondrial genome, where the proportion of LTR elements was unusually high, even surpassing that of *Cathaya argyrophylla*. This further highlights the differences in the patterns and mechanisms of repeat sequence formation between small and large mitochondrial genomes.

Further analysis of the relationship between repeat sequence fragment size and mitochondrial genome size revealed that, across all species, the majority of repeat sequences were small fragments between 0-100 bp, with their abundance decreasing as fragment size increased. However, in species with extremely large mitochondrial genomes (such as *Cathaya argyrophylla*, *Larix sibirica*, and *Picea sitchensis*), the proportion of repeat fragments longer than 200 bp increased with genome size. In contrast, this trend was not observed in smaller genomes, and in some cases, even showed the opposite pattern. This highlights a clear structural difference in the composition of repeat sequences between extremely large Pinaceae mitochondrial genomes and the smaller mitochondrial genomes of Pinaceae and other gymnosperms. It also suggests a distinct correlation between genome size and repeat sequence structure in extremely large mitochondrial genomes.

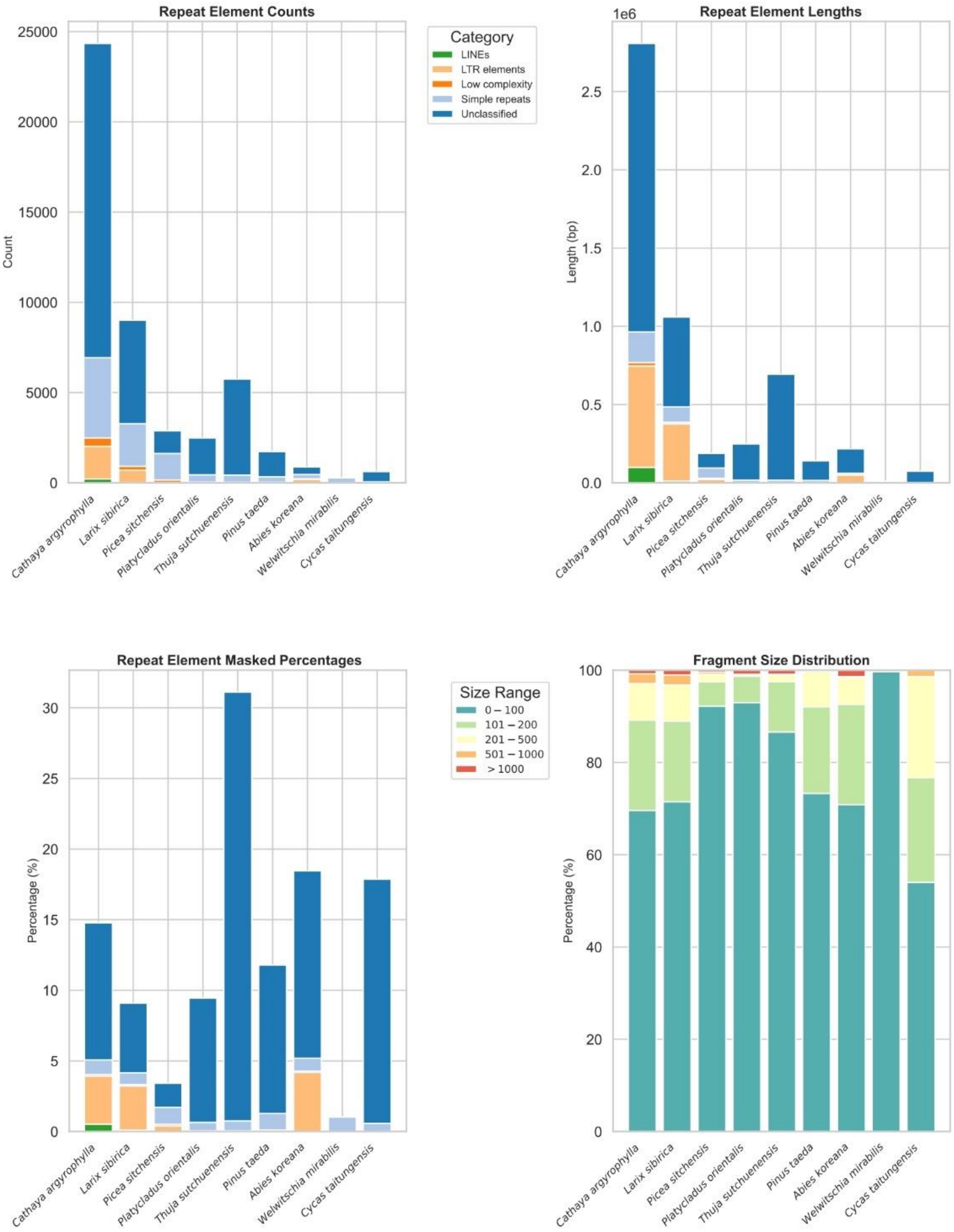

Figure 4. Comparison of repeat sequence types between *Cathaya argyrophylla* and other Pinaceae species.

Note: Species below the horizontal axis are arranged in descending order of mitochondrial genome size from left to right.

**Relationship between RNA Editing and Genome Size in Pinaceae**

Next, we evaluated RNA editing events in *Cathaya argyrophylla* using second-generation sequencing data from 10 leave samples of *Cathaya argyrophylla*, and compared these events with those in closely related Pinaceae species and other gymnosperms (Figure 5). Our analysis identified a total of 1,201 RNA editing sites in *Cathaya argyrophylla*, with the majority (1,172) being C-to-T edits. The number of RNA editing sites in *Cathaya argyrophylla* is the highest among large mitochondrial genomes of Pinaceae species. We observed a notable trend in plants with mitochondrial genomes larger than 6.17M (the size of *Picea smithiana*), where larger genomes tended to have more RNA editing sites. This trend was less clear in species with genomes between 3.85M and 6.17M, and did not apply to smaller species. For example, *Pinus taeda*, with a genome size of about 1M, exhibited as many as 1,179 editing sites. Moreover, while *Cathaya argyrophylla*, with the largest mitochondrial genome, has 1,201 RNA editing sites, this is not the highest among the gymnosperms compared in this study. *Ginkgo biloba*, with the smallest mitochondrial genome at 346,544 bp, had the most RNA editing sites, totaling 1,306. These findings suggest potential differences in the mechanisms and patterns of RNA editing between large and small mitochondrial genomes.

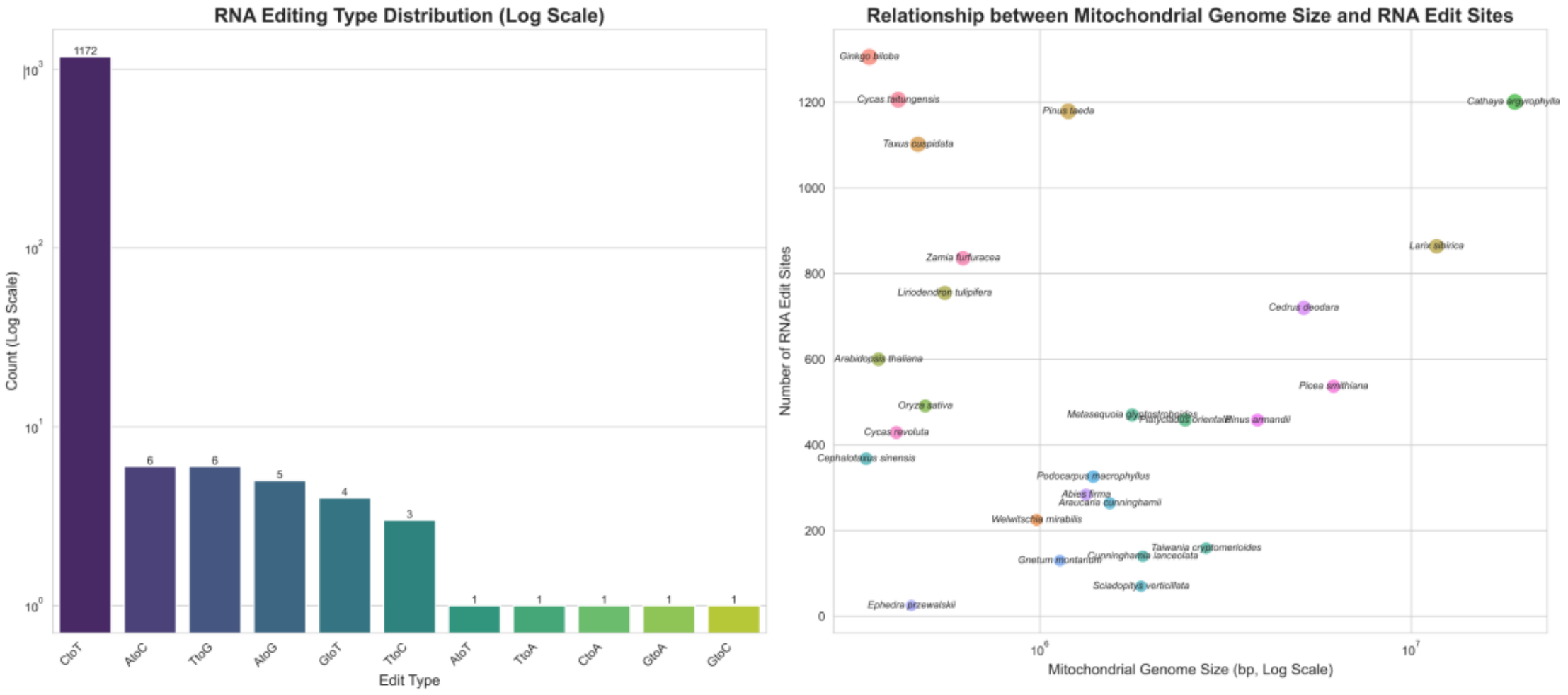

Figure 5. Comparative analysis of RNA editing sites between *Cathaya argyrophylla* and various other plants.

Note: The RNA editing sites for *Cathaya argyrophylla* were determined using second-generation sequencing data from this study. RNA editing site data for other species were obtained from the literature, including *Oryza sativa* [32], *Arabidopsis thaliana* [4], *Welwitschia mirabilis* [2], *Liriodendron tulipifera* [1], *Taxus cuspidata* [2], *Ginkgo biloba* [2], *Cycas taitungensis* [2], *Pinus taeda* [2], and *Larix sibirica* [17]. Data for the remaining species were sourced from reference [31].

**Relationship between Plastid-to-Mitochondrial DNA Transfers and Genome Size in Pinaceae**

To further compare the transfer of chloroplast/plastid sequences into the mitochondrial genomes of large Pinaceae species and other gymnosperms, we analyzed chloroplast-derived fragments (mtpts) in these genomes (Figure 6). The results showed that the number, total length, and coverage of mtpts in the chloroplast/plastid sequences were higher in the extremely large mitochondrial genomes of Pinaceae (*Cathaya argyrophylla*, *Larix sibirica*, and *Picea sitchensis*) compared to other gymnosperms. These three species consistently ranked among the top five for all three metrics, with *Cathaya argyrophylla* taking the top position, indicating a positive correlation between mitochondrial genome size and the transfer of chloroplast/plastid sequences.

A scatterplot and sequence transfer example diagram revealed a consistent trend across all three metrics: 1 million base pairs (1 Mbp) serve as a threshold for mitochondrial genome size (Figure 6). In genomes larger than this, mitochondrial genome size showed a positive correlation with the number, length, and coverage of mtpts, while this relationship did not hold in genomes smaller than 1Mbp. This suggests that mitochondrial genomes larger than 1 Mbp might experience different mechanisms or regulatory patterns for sequence exchange with the chloroplast genome compared to smaller gymnosperm mitochondrial genomes.

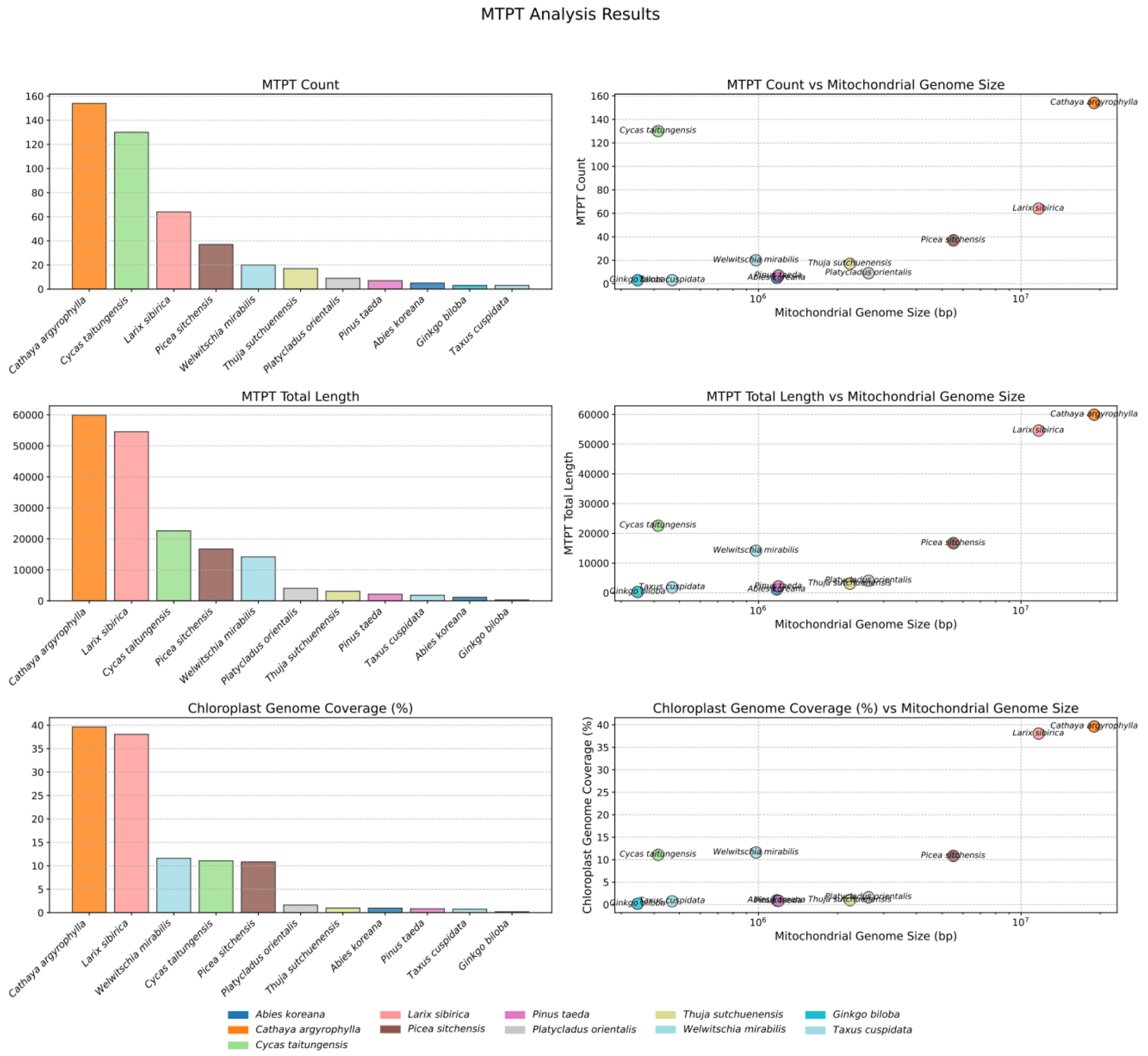

Figure 6. Comparative analysis of mtpt (chloroplast-derived sequences) between *Cathaya argyrophylla* and other gymnosperms.

Interestingly, the two largest mitochondrial genomes—*Cathaya argyrophylla* and Larix sibirica—had the highest mtpt coverage of chloroplast genomes, each around 40%, significantly surpassing other species (Figure 6). This indicates that species with extremely large mitochondrial genomes are more likely to incorporate chloroplast genome fragments and may share similar mechanisms and patterns of sequence exchange with chloroplast genomes.

We further analyzed the mtpt fragments in various mitochondrial genomes and specifically examined the transfer of chloroplast genes. The results showed that the mitochondrial genome of *Cathaya argyrophylla* had the highest number of transferred chloroplast genes, totaling 61, with 27 of them being fully transferred. In comparison, the second-largest mitochondrial genome, that of Siberian larch, had 48 transferred genes, also with 27 being complete transfers. The remaining mitochondrial genomes exhibited far fewer transferred chloroplast genes, all with fewer than 24 transfers, and a decreasing trend was observed as the genome size decreased. This indicates that larger mitochondrial genomes have a higher likelihood of acquiring transferred chloroplast genes, including full-length genes.

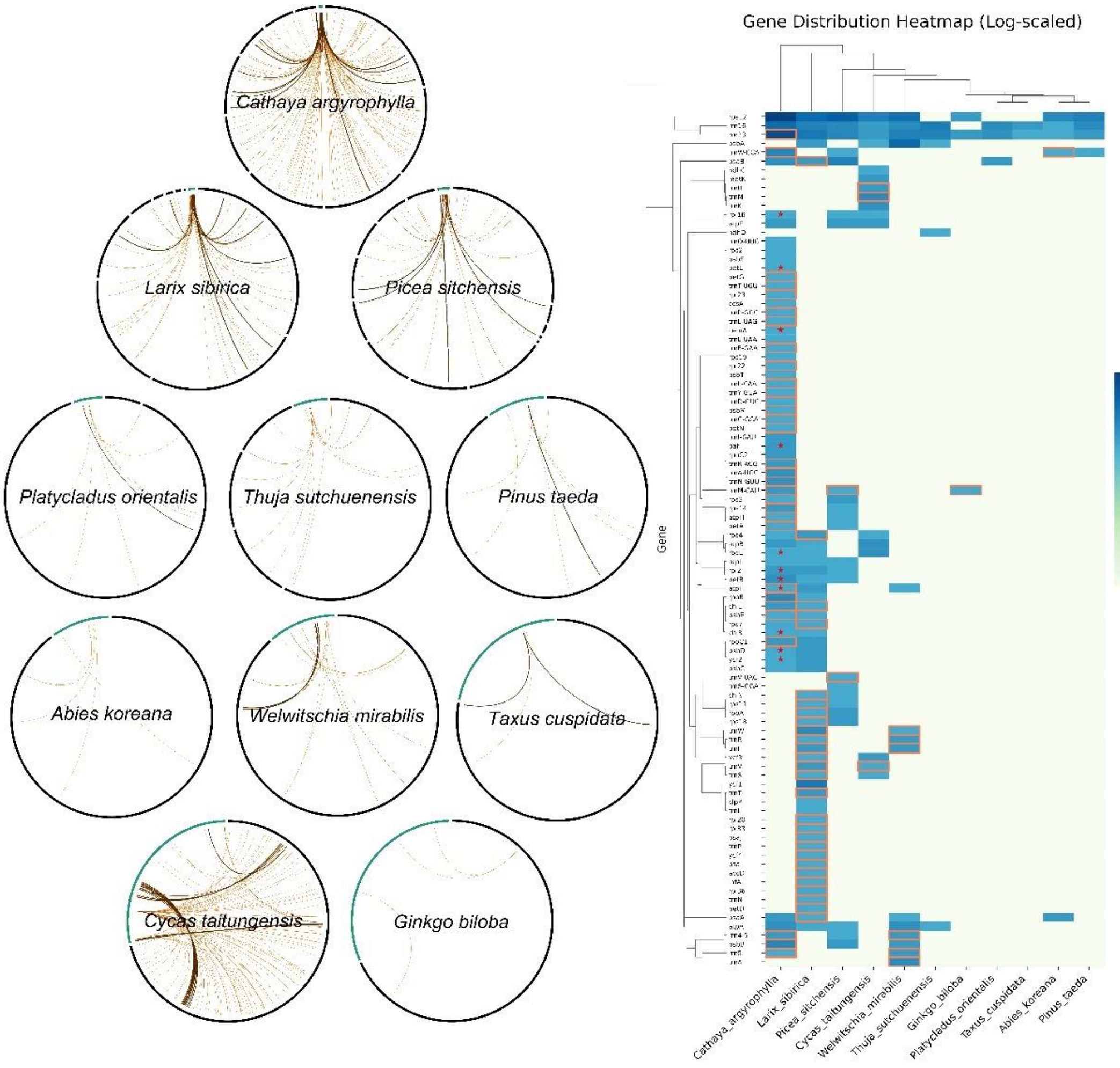

Figure 7. Comparison of chloroplast genes contained within mtpt sequences between *Cathaya argyrophylla* and other plants.

Note: The left panel shows the circos plots for each species' mtpt sequences. The outermost green ring represents the chloroplast genome, the black ring represents the mitochondrial genome contigs, and the brown lines indicate the connections between the mtpt sequences, where thicker lines represent larger mtpt segments. The right panel displays a heatmap comparing the chloroplast genes present in the mtpt sequences of *Cathaya argyrophylla* and other plants, with darker colors indicating higher quantities. Chloroplast genes surrounded by brown boxes indicate complete genes, and genes marked with a star symbol are chloroplast polymorphic genes identified in previous studies.

Upon closer inspection, it was also observed that 11 of the chloroplast genes transferred into the mitochondrial genome of *Cathaya argyrophylla* corresponded to previously identified polymorphic chloroplast genes across multiple populations of *Cathaya argyrophylla*, accounting for 57.9% of all 19 known polymorphic chloroplast genes. This suggests that these highly active polymorphic chloroplast genes are more likely to be transferred into the mitochondrial genome of *Cathaya argyrophylla*. Furthermore, a comparison between *Cathaya argyrophylla* and Siberian larch revealed that only a small number of the transferred genes were shared between the two species, indicating that the patterns of gene transfer between the mitochondria and chloroplast genomes can vary significantly between species.

**Analysis of Unique Mitochondrial-Nuclear DNA Exchange in Pinaceae**

Mitochondrial DNA fragments frequently transfer to the nuclear genome, forming numerous nuclear mitochondrial DNA sequences (NUMTs), which are often preserved and highly conserved among closely related species. To investigate the differences in sequence exchange between the mitochondrial and nuclear genomes of large Pinaceae species, such as *Cathaya argyrophylla*, compared to other gymnosperms and even angiosperms, we analyzed homologous fragments between the mitochondrial and nuclear genomes across Pinaceae, other gymnosperms, and angiosperms.

The results (Figure 8) revealed that the coverage of homologous fragments between the mitochondrial and nuclear genomes, specifically covering the mitochondrial genome, is relatively low in Pinaceae species. When using the nuclear genome of the same species as a reference, the highest coverage reached approximately 80%, with *Pinus taeda* showing the lowest coverage at only 40%. The coverage dropped even further when using the genomes of closely related species as references, particularly for extremely large mitochondrial genomes like those of *Cathaya argyrophylla* and *Larix sibirica*, where the coverage fell below 10%. However, a positive trend was observed as mitochondrial genome size increased. In Pinaceae species, the coverage of homologous fragments between the mitochondrial and nuclear genomes increased with mitochondrial genome size, from 50.04% in *Pinus taeda* (1.19M) to 78.94% in *Picea sitchensis* (5.5M) and 87.61% in *Larix sibirica* (11.7M).

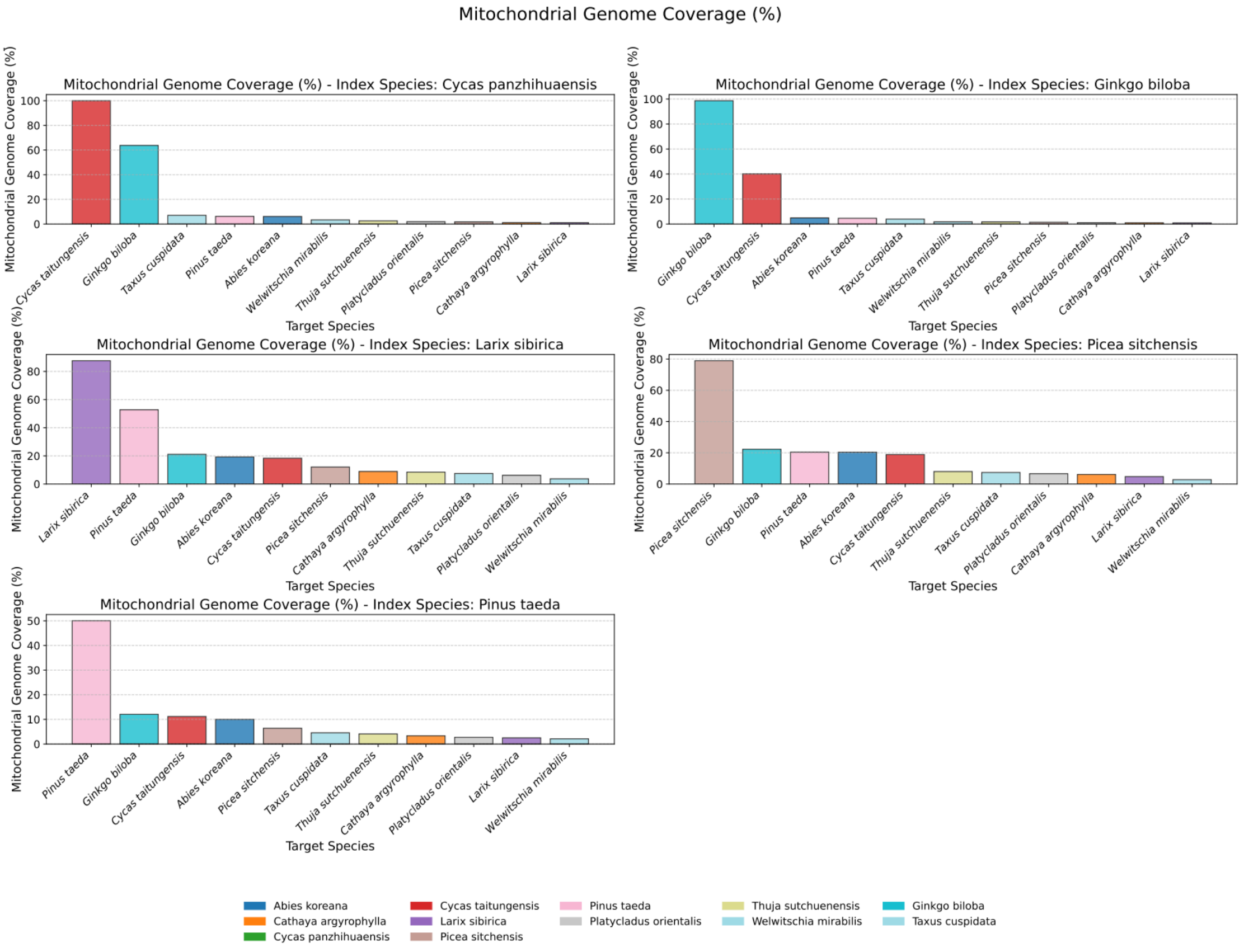

Figure 8. Nuclear Mitochondrial DNA Sequences (NUMTs) in Pinaceae.

Note: The figure shows NUMTs identified in *Cathaya argyrophylla* and other Pinaceae species, with different Pinaceae species used as index reference genomes. The vertical axis represents the coverage of these NUMTs in the mitochondrial genome.

In contrast, the situation was markedly different for small mitochondrial genomes in both angiosperms and gymnosperms. Angiosperms showed a higher degree of sequence exchange between the mitochondrial and nuclear genomes, with 100% coverage of homologous fragments covering the mitochondrial genome, when using the same species as a reference, and over 20% coverage even when using closely related species (Figure 9). A similar pattern was observed in gymnosperms with small mitochondrial genomes, such as *Ginkgo biloba* and *Cycas taitungensis*, where homologous fragment coverage between their mitochondrial and nuclear genomes exceeded 98% when using the same species as a reference and surpassed 40% when using each other's genomes as references.

These findings suggest that the mechanisms of sequence exchange between mitochondrial and nuclear genomes in Pinaceae species, particularly those with extremely large mitochondrial genomes, may differ significantly from those in gymnosperms with smaller mitochondrial genomes. This difference could reflect a complex and unique regulatory mechanism.

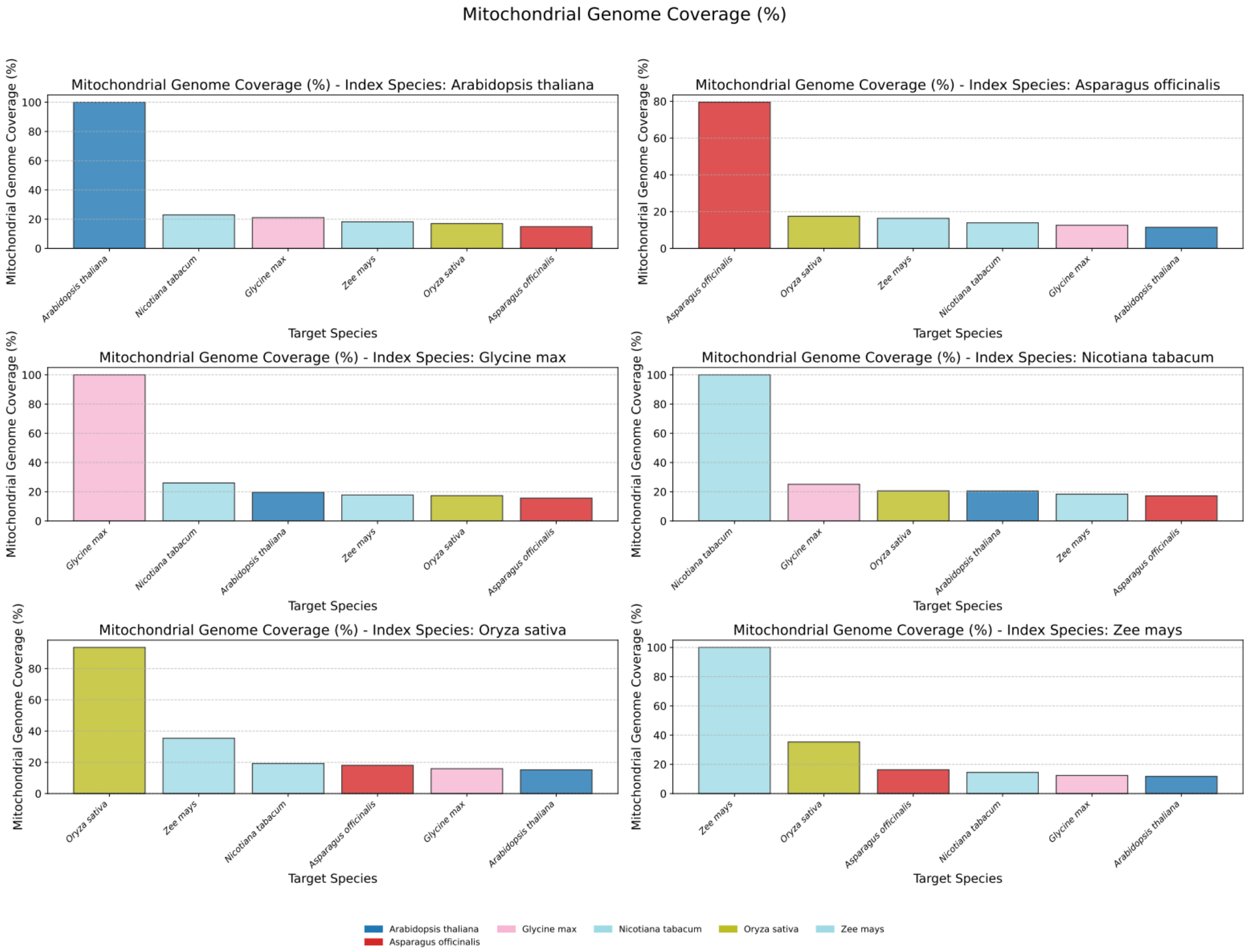

Figure 9. Nuclear Mitochondrial DNA Sequences (NUMTs) in Angiosperms.

Note: The figure shows NUMTs identified in angiosperms, with different Pinaceae species used as index reference genomes. The vertical axis represents the coverage of these NUMTs in the mitochondrial genome.

## Discussion

This study conducted a comprehensive analysis of the mitochondrial genome of *Cathaya argyrophylla*, uncovering its unique genomic features and comparing them with other Pinaceae species and gymnosperms. Through the investigation of repeat sequences, transposon activity, RNA editing events, chloroplast-derived fragments (mtpts), and the sequence homology between mitochondrial and nuclear genomes, we explored the common mechanisms and significance of extremely large mitochondrial genomes in Pinaceae.

Previous studies have consistently reported the large size and structural complexity of Pinaceae mitochondrial genomes, as exemplified by findings in *Larix sibirica* [17] and *Picea sitchensis* [9]. In this study, we discovered that the mitochondrial genome of *Cathaya argyrophylla* is the largest among all Pinaceae species, with a genome size of 18.99 Mb, surpassing the previously reported largest mitochondrial genome in *Larix sibirica* [17], setting a new record for mitochondrial genome size. This massive genome primarily consists of a multi-branched structure, and high-precision Nanopore sequencing and assembly revealed 6 linear contigs and 3 circular contigs. This is similar to the complex multi-contig structures found in *Larix sibirica* (9 contigs) and *Picea sitchensis* (13 contigs). However, compared to other extremely large mitochondrial genomes in Pinaceae, the genome size of *Cathaya argyrophylla* shows a significant increase, highlighting not only the shared characteristics of Pinaceae species but also the unique evolutionary features of *Cathaya argyrophylla*.

Synteny analysis plays a crucial role in understanding genome structure and evolution. By comparing syntenic relationships between genomes of different species, we can uncover evolutionary events such as genome rearrangements and infer phylogenetic relationships [4,11,12,33,34]. In this study, synteny analysis revealed that *Cathaya argyrophylla* shares numerous syntenic blocks with other Pinaceae species, with the number of syntenic blocks closely correlating with phylogenetic relatedness. This indicates that the mitochondrial genomes of Pinaceae have maintained a certain level of conservation throughout evolution. However, we also observed that the number of syntenic blocks is influenced by mitochondrial genome size. Species with larger mitochondrial genomes tend to share more syntenic blocks, suggesting that genome expansion may facilitate sequence sharing and recombination. This finding supports the role of genome size in influencing genome structure and evolution, indicating that larger genomes may have undergone more frequent recombination and rearrangement events.

Repeat sequence analysis is a key tool for understanding genome expansion and increasing complexity. The accumulation of repeat sequences can lead to genome size enlargement and influence genome stability and functionality [4,6,14,35,36]. In this study, we found that while *Cathaya argyrophylla* and Larix sibirica have the highest total length of repeat sequences and the highest total length of their genomes covered by repeats, their overall repeat sequence coverage rate is not the highest. This suggests that even extremely large mitochondrial genomes are not predominantly composed of repeat sequences. Additionally, in extremely large mitochondrial genomes, the proportion of repeat sequences longer than 200 bp increases with genome size, a trend that is not observed in smaller genomes. This implies that extremely large mitochondrial genomes may have unique mechanisms for the formation and accumulation of repeat sequences.

Transposon activity plays a crucial role in genome evolution and structural variation. The insertion and amplification of transposons can lead to changes in genome size and gene content[3,4,37,38]. Previous studies have shown that plant mitochondrial genomes contain relatively few transposon types and numbers, particularly the rare LINEs and LTR elements[17,39,40], and our research confirms this observation. However, our study further reveals an increase in the presence of LINEs and LTR elements in extremely large mitochondrial genomes. We found that as mitochondrial genome size increases, the proportion of LINEs and LTR elements also rises, suggesting that transposon activity may have played a key role in the expansion of extremely large mitochondrial genomes. However, In contrast, this trend was not observed in smaller mitochondrial genomes, and in some cases, an unusually high proportion of LTR elements was found (such as in *Abies koreana*), indicating that different regulatory mechanisms may govern transposon activity in genomes of different sizes. Despite the potential link between extremely large mitochondrial genome size and the increased activity of certain transposon types like LINEs and LTR elements, repeat sequences may only partially explain the massive size of large Pinaceae mitochondrial genomes. For instance, even in *Cathaya argyrophylla*, which has the highest mitochondrial genome size, these elements account for only 14.78% of the genome, suggesting that other factors contribute to the substantial genome size.

The transfer of chloroplast-derived fragments (mtpts) is a key pathway for inter-organellar genome exchange. The integration of mtpts can enlarge mitochondrial genomes and provide new sequences for genome recombination [3,4,13,36,41]. Our study shows that *Cathaya argyrophylla*, with the largest mitochondrial genome, has the highest number, total length, and coverage of mtpts in its mitochondrial genome, and for species with mitochondrial genome sizes exceeding 1 Mbp, all mtpt metrics are positively correlated with genome size. Notably, the mtpt coverage in *Cathaya argyrophylla* and *Larix sibirica* extends to approximately 40% of the chloroplast genome, far exceeding that of other species. This suggests that larger mitochondrial genomes may be more receptive to incorporating chloroplast sequences, thereby promoting genome expansion and increased complexity. Furthermore, a significant proportion of the transferred chloroplast genes were previously identified as polymorphic chloroplast genes in *Cathaya argyrophylla* [18], confirming their activity and laying a foundation for further functional studies of these genes.

The exchange between mitochondrial and nuclear genomes is crucial for understanding genome evolution and organelle-nucleus interactions. In many plants, mitochondrial DNA fragments can be transferred to the nuclear genome, forming nuclear mitochondrial DNA sequences (NUMTs) that influence the structure and function of the nuclear genome [10,37,42,43]. However, we found that the sequence homology between the mitochondrial and nuclear genomes in Pinaceae species is relatively low, contrasting with findings in angiosperms and other gymnosperms. There are two possible explanations for this phenomenon: (1) Pinaceae species may have reduced mitochondrial-nuclear genome exchange, resulting in lower homology coverage; (2) the high variability and complexity of mitochondrial genome sequences in Pinaceae could complicate sequence comparisons, leading to lower coverage when closely related species are used as references. Additionally, as the mitochondrial genome size increases in Pinaceae species, the coverage of homologous fragments within the mitochondrial genome itself also increases when using the species' own nuclear genome as a reference. This may indicate that the expansion of mitochondrial genomes could enhance the potential or activity of sequence exchange with the nuclear genome.

Based on the results, we hypothesize that the formation of large mitochondrial genomes in Pinaceae species is likely the result of multiple contributing factors. While the increase in repeat sequences and transposon activity may have facilitated genome expansion, they are not the primary driving forces. Instead, the extensive incorporation of plastid sequences appears to play a more critical role in genome enlargement. Additionally, the increased number of RNA editing events may be linked to the heightened complexity of the genome, reflecting the functional regulatory demands of large mitochondrial genomes. Furthermore, the relatively limited exchange between the mitochondrial and nuclear genomes in Pinaceae suggests that unique mechanisms may govern their interaction. These factors exhibit different patterns in large mitochondrial genomes compared to smaller ones, indicating that Pinaceae species might possess a distinct evolutionary mechanism for mitochondrial genome development. These findings are important for understanding the structure, function, and evolution of plant mitochondrial genomes. The uniqueness of large Pinaceae mitochondrial genomes suggests the presence of specialized regulatory mechanisms that could influence genome stability, energy metabolism, and adaptive evolution. Future research should further investigate these mechanisms, particularly the roles of plastid sequence transfer and transposon activity in genome expansion, as well as the processes governing sequence exchange between large mitochondrial genomes and the nuclear genome.

In this study, *Cathaya argyrophylla* set a new record for mitochondrial genome size, deepening our understanding of large mitochondrial genomes in Pinaceae. These findings are not only significant for research on the structure of plant mitochondrial genomes, but they also provide new evidence for understanding the evolution and diversity of plant genomes. Future studies should integrate functional genomics and molecular biology approaches to further investigate these issues. Additionally, the conservation and utilization of *Cathaya argyrophylla* could benefit from these foundational studies, offering key insights into genetic diversity and evolutionary adaptability.

## References

1 Edera, A. A., Gandini, C. L. & Sanchez-Puerta, M. V. Towards a comprehensive picture of C-to-U RNA editing sites in angiosperm mitochondria. *Plant Molecular Biology* **97**, 215-231, doi:10.1007/s11103-018-0734-9 (2018).

2 Kan, S.-L., Shen, T.-T., Gong, P., Ran, J.-H. & Wang, X.-Q. The complete mitochondrial genome of Taxus cuspidata (Taxaceae): eight protein-coding genes have transferred to the nuclear genome. *BMC Evolutionary Biology* **20**, 10, doi:10.1186/s12862-020-1582-1 (2020).

3 Mower, J. P. Variation in protein gene and intron content among land plant mitogenomes. *Mitochondrion* **53**, 203-213, doi:https://doi.org/10.1016/j.mito.2020.06.002 (2020).

4 Small, I. D., Schallenberg-Rüdinger, M., Takenaka, M., Mireau, H. & Ostersetzer-Biran, O. J. T. P. J. Plant organellar RNA editing: what 30 years of research has revealed. **101**, 1040-1056 (2020).

5 Wang, S. *et al.* Evolution and Diversification of Kiwifruit Mitogenomes through Extensive Whole-Genome Rearrangement and Mosaic Loss of Intergenic Sequences in a Highly Variable Region. *Genome Biology and Evolution* **11**, 1192-1206, doi:10.1093/gbe/evz063 (2019).

6 Yu, R. *et al.* The minicircular and extremely heteroplasmic mitogenome of the holoparasitic plant <em>Rhopalocnemis phalloides</em>. *Current Biology* **32**, 470-479.e475, doi:10.1016/j.cub.2021.11.053 (2022).

7 Choi, K.-S. & Park, S. J. I. j. o. m. s. Complete plastid and mitochondrial genomes of Aeginetia indica reveal intracellular gene transfer (IGT), horizontal gene transfer (HGT), and cytoplasmic male sterility (CMS). **22**, 6143 (2021).

8 Yu, R., Sun, C., Liu, Y. & Zhou, R. J. P. Shifts from cis-to trans-splicing of five mitochondrial introns in Tolypanthus maclurei. **9**, e12260 (2021).

9 Jackman, S. D. *et al.* Complete Mitochondrial Genome of a Gymnosperm, Sitka Spruce (Picea sitchensis), Indicates a Complex Physical Structure. *Genome Biology and Evolution* **12**, 1174-1179, doi:10.1093/gbe/evaa108 (2020).

10 Jiang, M., Ni, Y., Li, J. & Liu, C. Characterisation of the complete mitochondrial genome of Taraxacum mongolicum revealed five repeat-mediated recombinations. *Plant Cell Reports* **42**, 775-789, doi:10.1007/s00299-023-02994-y (2023).

11 Sloan, D. B., Wu, Z. & Sharbrough, J. Correction of Persistent Errors in Arabidopsis Reference Mitochondrial Genomes. *The Plant Cell* **30**, 525-527, doi:10.1105/tpc.18.00024 (2018).

12 Sang, S.-F. *et al.* Organelle genome composition and candidate gene identification for Nsa cytoplasmic male sterility in Brassica napus. *BMC Genomics* **20**, 813, doi:10.1186/s12864-019-6187-y (2019).

13 Sullivan, A. R. *et al.* The Mitogenome of Norway Spruce and a Reappraisal of Mitochondrial Recombination in Plants. *Genome Biology and Evolution* **12**, 3586-3598, doi:10.1093/gbe/evz263 (2020).

14 Chevigny, N., Schatz-Daas, D., Lotfi, F. & Gualberto, J. M. J. I. J. o. M. S. DNA repair and the stability of the plant mitochondrial genome. **21**, 328 (2020).

15 Cheng, Y. *et al.* Assembly and comparative analysis of the complete mitochondrial genome of Suaeda glauca. *BMC Genomics* **22**, 167, doi:10.1186/s12864-021-07490-9 (2021).

16 Kozik, A. *et al.* The alternative reality of plant mitochondrial DNA: One ring does not rule them all. **15**, e1008373 (2019).

17 Putintseva, Y. A. *et al.* Siberian larch (Larix sibirica Ledeb.) mitochondrial genome assembled using both short and long nucleotide sequence reads is currently the largest known mitogenome. *BMC Genomics* **21**, 654, doi:10.1186/s12864-020-07061-4 (2020).

18 Huang, K., Mo, P., Deng, A., Xie, P. & Wang, Y. J. G. Differences in the Chloroplast Genome and Its Regulatory Network among *Cathaya argyrophylla* Populations from Different Locations in China. **13**, 1963 (2022).

19 Xie, P. *et al.* The diversity and abundance of bacterial and fungal communities in the rhizosphere of *Cathaya argyrophylla* are affected by soil physicochemical properties. **14**, 1111087 (2023).

20 Wang, H. & Ge, S. Phylogeography of the endangered *Cathaya argyrophylla* (Pinaceae) inferred from sequence variation of mitochondrial and nuclear DNA. *Molecular Ecology* **15**, 4109-4122 (2006).

21 Wynn, E. L. & Christensen, A. C. Repeats of Unusual Size in Plant Mitochondrial Genomes: Identification, Incidence and Evolution. *G3 Genes|Genomes|Genetics* **9**, 549-559, doi:10.1534/g3.118.200948 (2019).

22 Kolmogorov, M., Yuan, J., Lin, Y. & Pevzner, P. A. Assembly of long, error-prone reads using repeat graphs. *Nature Biotechnology* **37**, 540-546, doi:10.1038/s41587-019-0072-8 (2019).

23 Li, H. & Durbin, R. Fast and accurate short read alignment with Burrows–Wheeler transform. *Bioinformatics* **25**, 1754-1760, doi:10.1093/bioinformatics/btp324 (2009).
24 Wick, R. R., Judd, L. M., Gorrie, C. L. & Holt, K. E. J. P. c. b. Unicycler: resolving bacterial genome assemblies from short and long sequencing reads. **13**, e1005595 (2017).
25 Wick, R. R., Schultz, M. B., Zobel, J. & Holt, K. E. Bandage: interactive visualization of de novo genome assemblies. *Bioinformatics* **31**, 3350-3352, doi:10.1093/bioinformatics/btv383 (2015).
26 Dunn, N. A. *et al.* Apollo: democratizing genome annotation. **15**, e1006790 (2019).
27 Wang, Y. *et al.* MCScanX: a toolkit for detection and evolutionary analysis of gene synteny and collinearity. *Nucleic Acids Research* **40**, e49-e49, doi:10.1093/nar/gkr1293 (2012).
28 Abrusán, G., Grundmann, N., DeMester, L. & Makalowski, W. TEclass—a tool for automated classification of unknown eukaryotic transposable elements. *Bioinformatics* **25**, 1329-1330, doi:10.1093/bioinformatics/btp084 (2009).
29 Chen, Y., Ye, W., Zhang, Y. & Xu, Y. High speed BLASTN: an accelerated MegaBLAST search tool. *Nucleic Acids Research* **43**, 7762-7768, doi:10.1093/nar/gkv784 (2015).
30 Chen, C. *et al.* TBtools: An Integrative Toolkit Developed for Interactive Analyses of Big Biological Data. *Molecular Plant* **13**, 1194-1202, doi:10.1016/j.molp.2020.06.009 (2020).
31 Kan, S.-L., Shen, T.-T., Ran, J.-H. & Wang, X.-Q. Both Conifer II and Gnetales are characterized by a high frequency of ancient mitochondrial gene transfer to the nuclear genome. *BMC Biology* **19**, 146, doi:10.1186/s12915-021-01096-z (2021).
32 Notsu, Y. *et al.* The complete sequence of the rice (Oryza sativa L.) mitochondrial genome: frequent DNA sequence acquisition and loss during the evolution of flowering plants. *Molecular Genetics and Genomics* **268**, 434-445, doi:10.1007/s00438-002-0767-1 (2002).
33 Kovar, L. *et al.* PacBio-Based Mitochondrial Genome Assembly of Leucaena trichandra (Leguminosae) and an Intrageneric Assessment of Mitochondrial RNA Editing. *Genome Biology and Evolution* **10**, 2501-2517, doi:10.1093/gbe/evy179 (2018).
34 Rawal, H. C., Kumar, P. M., Bera, B., Singh, N. K. & Mondal, T. K. Decoding and analysis of organelle genomes of Indian tea (Camellia assamica) for phylogenetic confirmation. *Genomics* **112**, 659-668, doi:https://doi.org/10.1016/j.ygeno.2019.04.018 (2020).
35 Dong, S. *et al.* The complete mitochondrial genome of the early flowering plant Nymphaea colorata is highly repetitive with low recombination. *BMC Genomics* **19**, 614, doi:10.1186/s12864-018-4991-4 (2018).
36 Martins, G., Balbino, E., Marques, A. & Almeida, C. Complete mitochondrial genomes of the Spondias tuberosa Arr. Cam and Spondias mombin L. reveal highly repetitive DNA sequences. *Gene* **720**, 144026, doi:https://doi.org/10.1016/j.gene.2019.144026 (2019).
37 Wang, X. *et al.* Comprehensive analysis of complete mitochondrial genome of Sapindus mukorossi Gaertn.: an important industrial oil tree species in China. *Industrial Crops and Products* **174**, 114210, doi:https://doi.org/10.1016/j.indcrop.2021.114210 (2021).
38 Li, J. *et al.* Assembly of the complete mitochondrial genome of an endemic plant, Scutellaria tsinyunensis, revealed the existence of two conformations generated by a repeat-mediated recombination. *Planta* **254**, 36, doi:10.1007/s00425-021-03684-3 (2021).
39 Clifton, S. W. *et al.* Sequence and Comparative Analysis of the Maize NB Mitochondrial Genome. *Plant Physiology* **136**, 3486-3503, doi:10.1104/pp.104.044602 (2004).
40 Knoop, V. *et al.* copia-, gypsy- and LINE-Like Retrotransposon Fragments in the Mitochondrial

Genome of Arabidopsis thaliana. *Genetics* **142**, 579-585, doi:10.1093/genetics/142.2.579 (1996).

41 Xia, C. *et al.* Complete mitochondrial genome of Thuja sutchuenensis and its implications on evolutionary analysis of complex mitogenome architecture in Cupressaceae. *BMC Plant Biology* **23**, 84, doi:10.1186/s12870-023-04054-9 (2023).

42 Richly, E. & Leister, D. NUMTs in Sequenced Eukaryotic Genomes. *Molecular Biology and Evolution* **21**, 1081-1084, doi:10.1093/molbev/msh110 (2004).

43 Hazkani-Covo, E., Zeller, R. M. & Martin, W. J. P. g. Molecular poltergeists: mitochondrial DNA copies (numts) in sequenced nuclear genomes. **6**, e1000834 (2010).